\documentclass[showpacs,amsmath,amssymb,prl,twocolumn]{revtex4-1}
\usepackage{graphicx}
\usepackage{color}

\begin{document}
\title{A self-organized critical model of rearranging hydrogen-bonded
network in ice}
\author{ Tridib Sadhu}
\affiliation{
Department of Theoretical Physics, \\ 
Tata Institute of Fundamental Research, \\ 
Homi Bhabha Road, Mumbai 400 005, India.}
\begin{abstract}
A dynamical four-vertex model of rearranging hydrogen-bonded network in
ice, with a dynamics similar to the Grotthuss mechanism, is proposed.
The model qualitatively explains the unusually high proton-conductivity
in ice. It also serves as an interesting example of self-organized
criticality with a non-conservative dynamics. The model is solved on a
linear chain and its steady state is determined. Using numerical
simulations, the model is further studied on a square lattice, and it
is observed that it relaxes by three different avalanches, each of
which follows a simple finite-size scaling.
\end{abstract}
\maketitle

An isolated water molecule can be said to be well understood because its
properties, as observed in experiments, can be described by the first
principles
quantum mechanical calculations. However, when a large number of
these molecules form liquid-water or ice, our ability to predict their
properties becomes limited \cite{ball}. In fact, water has a large number of
anomalous properties \cite{63}, amongst which the unusually high
mobility of $\textrm{H}^{+}$ has attracted a lot of attention. The observed
value of the mobility in bulk ice \cite{expt} or water-filled narrow
pores \cite{onemd} are
comparable to the electronic mobilities in some semiconductors. Conduction of
$\textrm{H}^{+}$ is also fundamental to a myriad of processes ranging from ATP synthesis to electrical power
generation in hydrogen fuel cells.

\begin{figure}
\includegraphics[scale=0.20,angle=0]{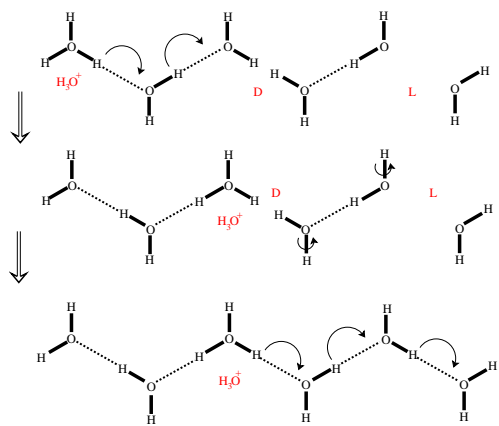}
\caption{(Color online) A schematic of $\textrm{H}^{+}$ transport by the Grotthuss
mechanism. The hydronium ion and the two types of broken bonds, D and L
are shown by red letters. The hopping of $\textrm{H}^{+}$ and the rotation of
molecules around an O-H axis are indicated
by arrows. }
\label{fig:ornt}
\end{figure}

The generally accepted idea is that the hydrogen-bonded network of water molecules
provides conduction pathways to $\textrm{H}^{+}$, which migrates via Grotthuss mechanism
\cite{gm}. In this, an extra $\textrm{H}^{+}$ in the network appears as a
hydronium ion ($\textrm{H}_{\textrm{3}}\textrm{O}^+$), which
translocates by hopping of $\textrm{H}^{+}$ across the hydrogen bonds (Fig. \ref{fig:ornt}). This movement is hindered by the broken hydrogen bonds in
the network (now on referred to as defects), which themselves migrate by
rotation of the water molecules. The rotation of the molecules is slower than the hopping of
$\textrm{H}^{+}$ and is therefore the rate
limiting step \cite{gm}. However, a comparison of the mobility of
$\textrm{H}^{+}$, determined experimentally, with that of the defects,
estimated assuming a
simple diffusion, shows that the
former is much higher at low temperatures \cite{agmon}.
We show that this problem can be qualitatively resolved if one considers
a long-range correlation in the hydrogen-bonded network. While presence of
this correlation has been experimentally observed and its
importance envisioned \cite{lrcexpt}, a clear understanding is still
missing. In this letter, we propose a model of rearranging
hydrogen-bonded network, with a Grotthuss-type dynamics, which
exhibits a long-range correlation in its steady state. This correlation in
the network increases the
mobility of the defects, and a long distance transport of these occurs
with a non-zero probability.
We solve this model on a linear chain and determine its steady state.
Further, using numerical simulations on a square lattice, we show
that, in the steady state, the
probability distribution of long distance transport of the defects can
be described by a linear combination of simple scaling forms.

The model is interesting \textit{per se}, as it provides a theoretical example of Self-Organized Criticality (SOC) with
non-conservative dynamics. The existence of non-conservative models of SOC has been long debated.
The much studied models, like the forest fire model \cite{forest},
the Olami-Feder-Christensen model in the non-conservative limit
\cite{ofc} and their variants are known to be
non-critical \cite{bonachela}. The non-conservative sandpile model
studied by Pruessner and Jensen \cite{pruessner} is critical in
the infinite volume limit, but lacks a well defined finite-size
scaling, unless a control parameter is fine-tuned \cite{bonachela}.
The model we propose does not have any conserved
quantities. When slowly
driven, it exhibits a scale invariance,
following a finite-size scaling, without any apparent fine tuning of
parameters. As the model shares some features with the well known ice
model \cite{lieb} of equilibrium statistical mechanics and also with
the sandpile model of SOC \cite{dhar06},
we call it \textit{the icepile model}.
\begin{figure}
\begin{center}
\includegraphics[scale=0.20,angle=0]{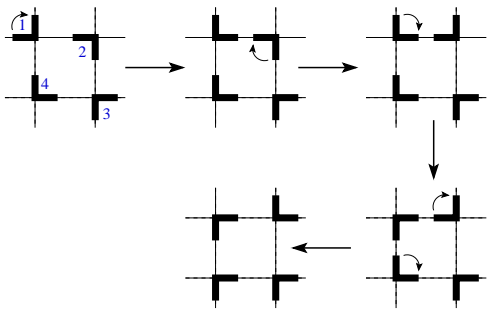}
\caption{An illustration of the relaxation, started by
rotating the molecule at the top-left site of a $2\times 2$ lattice. The covalent O-H bonds are
denoted by solid lines, and the rotation of molecules are
indicated by arrows. The four orientations of a molecule are
denoted by $1,2,3$ and $4$.}
\label{fig:grnd}
\end{center}
\end{figure}

We define the model on a square lattice with open boundaries (Fig. \ref{fig:grnd}).
Every site on the lattice is occupied by an oxygen atom. Two hydrogen atoms
are connected to each site by covalent bonds and they are always at
right angles to each other. This is a special case of the
standard ice model, with only four out of the six
orientations allowed, \textit{i.e.}, the orientations where two
hydrogens are diametrically opposite to each other, are not allowed.

Any orientational arrangement of the molecules may
have empty edges or edges occupied by two hydrogen atoms. These are known as
L- and D-defects \cite{gm}, respectively. Hydrogens in
the neighboring molecules interact via repulsive
electrostatic interaction, and the D-defects cost energy.
Let the energy be $U$ for a D-defect, and zero for the singly-
or un-occupied edges.
Further, to specify the orientations of the water molecules, we introduce a set of variables
$\tau_{i,j}$ such that $\tau_{i,j}=1$, when a hydrogen atom from the
$i$th site is along the $i$-$j$ edge, and $\tau_{i,j}=0$, otherwise.
The energy of a configuration is then described by an effective Hamiltonian
\begin{equation}
\mathcal{H}=\frac{U}{2}\sum_{<i,j>} \tau_{i,j}\tau_{j,i},
\label{hamiltonian}
\end{equation}
where the angular brackets indicate that $i$ and $j$ are nearest
neighbors, and $\tau_{i,j}=0$ for sites outside the boundary.
The equilibrium properties corresponding to this Hamiltonian are easy
to determine, if we
represent each orientation of a water molecule by a pair of
Ising-spins, taking up or down and left or
right orientations. This breaks the system into mutually uncoupled horizontal
and vertical linear chains of Ising-spins, which is exactly
solvable.

As mentioned before, the translocation of the defects
determines the effective mobility of $\textrm{H}^{+}$, and this is what we study in
the non-equilibrium steady state of our model. We assume the configurations with D-defects as unstable.
In an ice crystal, the defects are formed mostly by thermal
excitations and their diffusion takes place by the reorientation
of water molecules \cite{gm}. We generate the defects by
forcibly reorienting a randomly chosen molecule by $90^{\circ}$ clockwise
rotation, calling it an \textit{external drive}. If this creates a
D-defect, we use the following odd-even update rule to relax the
configuration (see Fig.
\ref{fig:grnd}): We divide the lattice into two sub-lattices
according to the parity of $x+y$, where $x$ and $y$ are the Cartesian
coordinates of a site. We pick one sub-lattice and select all sites
on it contributing to a D-defect and
simultaneously rotate the corresponding water molecules by
$90^{\circ}$ clockwise. This
may create or annihilate D-defects and/or move them by one lattice unit.
We do the same for the other sub-lattice and repeat until
the system reaches a stable configuration. At this stage we again disturb
the system by forcibly rotating another randomly chosen molecule and
repeat.

We refer to a complete relaxation process, starting from the driving to
reaching a stable configuration, as an avalanche. 
The number of D-defects is not conserved in an avalanche.
The reorientation of the molecules in a relaxation step may create more
D-defects which move on the lattice until they meet an empty edge or the
lattice boundary, where they annihilate. Unlike the conserved sandpile models, the
activity during a relaxation is not coupled to any conserved
quantities.

There are a large number of stable configurations, all of which appear
with a non-zero probability in the steady state.
This can be verified from the fact
that the configuration with all water molecules in the same
orientation can be reached from any stable configuration, and
\textit{vice versa}.
In a stable configuration, each row and column in the lattice has at most one
empty edge whose positions uniquely determine the configuration.
For each row with $L$ sites, there are $L+1$ ways of placing the empty edge (we
consider the configuration with no empty edges as the one
with empty edges outside the lattice). Then for an $L\times M$
lattice, there are
\begin{equation}
\Omega_{square}=\left( L+1 \right)^{M}\left( M+1 \right)^{L}
\end{equation}
ways of placing the empty edges, thus, so many stable configurations.

We first study an $L\times 1$ chain and show that all the $2^{L}\left( L+1\right)$
stable configurations are equally probable in the steady state.
Let $| C \rangle$ be a stable configuration. Define an operator $a_{i}$ by the relation
$a_{i}|C\rangle=|C'\rangle$, where $|C'\rangle$ is the
configuration reached by driving at site $i$ of the chain in
configuration
$| C \rangle$, and relaxing. The D-defect created by the driving, moves
toward the empty edge and vanishes by annihilating with it. It is
always the molecule at the site next to the empty edge, which rotates
last. For the present case, let that site be $j$.
We define an operator $R$, corresponding to the reflection of the system
across a line parallel to the chain, which maps any allowed configuration to another unique allowed
configuration. Then it can be shown that there exists an avalanche corresponding to the equation $a_{j}R|C'\rangle=R |C\rangle$.
Similar argument applies for all the driving operators, implying that
they all have an inverse. Thus the Markov evolution satisfies the pairwise balance
\cite{pair}, and hence the steady state is equally probable.

Let $s$ be the total number of $90^{\circ}$ rotations of the water molecules in an
avalanche, and $P\left( s,L \right)$ be the probability of such events
on a linear chain of $L$ sites.
As noted, the D-defect generated by the
driving moves toward the empty edge, and vanishes by annihilating with it.
All the sites in its path rotate at least once. If the driving site is on the right of
the empty edge, then, before relaxation, the sites in between are either in
state $3$ or $4$ (Fig. \ref{fig:grnd}). For the configurations where all these sites are in state
$3$, the corresponding water molecules rotate just once. If any of
these sites are in state $4$, then, each of them individually increases the total number of
rotations by $2$. Thus, effectively, the sites in state $3$ and
$4$ contribute $s=1$ and $s=3$, respectively. A similar argument applies for the driving
sites on the left of the empty
edge. From this it is easy to show that
the probability
\begin{equation}
P\left( s+1, L \right)=\sum_{k=0}^{\lfloor s/3 \rfloor} {s-2k \choose k} 2^{2k-s}\frac{L-\left( s-2k \right)}{L\left( L+1 \right)}+\frac{1}{2}\delta_{s,0},
\label{distr}
\end{equation}
where $\lfloor x \rfloor$ is the largest integer less than or equal to
$x$ and $\delta_{i,j}$ is the Kronecker delta, which corresponds to the
fact that half of the times the driving does not create any D-defect.
For a large $s$, the expression reduces to a
simple scaling form
\begin{equation}
P\left( s, L\right)\simeq L^{-1} g \left( s/L \right),
\label{scalone}
\end{equation}
where $g\left( x \right)=2^{-1}\left( 1-x/2 \right)$.

Let us now consider an $L \times L$ lattice. Our numerical studies show
that, in the steady
state, the empty edges are more likely to be found near the center of
the lattice. This implies that the stable configurations in the steady
state are not equally probable.

\begin{figure}
\includegraphics[scale=0.12,angle=0]{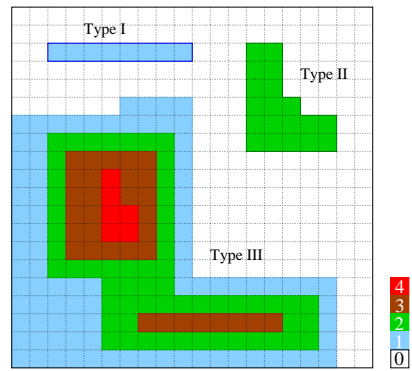}
\caption{(Color online) A schematic of the affected regions
in the three types of avalanches on a $20\times20$ lattice. Each
colored box is a unit cell of the lattice. Different colors represent,
different number of rotations per site.}
\label{fig:avalanche}
\end{figure}
On this lattice there are three types of avalanches. A schematic of
these avalanches are shown in Fig. \ref{fig:avalanche}.
In type I, the activity is confined along a single row or a column.
In the other two types, the activity spreads on a two-dimensional area.
In one of these, the activity moves like a single wave crossing the sites
in its path only once, and the water molecules rotate less than five times. This is the type
II avalanche.
In the other, the waves move back and forth on the lattice, until the
system reaches a stable configuration. This is the type III avalanche. The number of
rotations increases for the sites which are away from the boundary of the
affected region, and has a maximum value of the order $L$.
Then the cutoff in the total number of rotations, $s$, in an avalanche, scales as
$L$, $L^2$ and $L^{3}$ for the three types of avalanches, respectively.
In such a case, the probability distribution is expected to have a scaling form
\begin{equation}
P(s,L)=\frac{1}{L^{\alpha}}f\left( \frac{s}{L}
\right)+\frac{1}{L^{\beta}}g\left( \frac{s}{L^{2}}
\right)+\frac{1}{L^{\nu}}h\left( \frac{s}{L^{3}} \right),
\label{linscal}
\end{equation}
with the scaling functions $f,g$ and $h$, which vanish as their argument
approaches $\mathcal{O}\left( 1 \right)$. A similar linear combination of simple scaling forms
was found in a few other models \cite{ali,priezzev}.

\begin{figure}
\includegraphics[scale=0.65,angle=0]{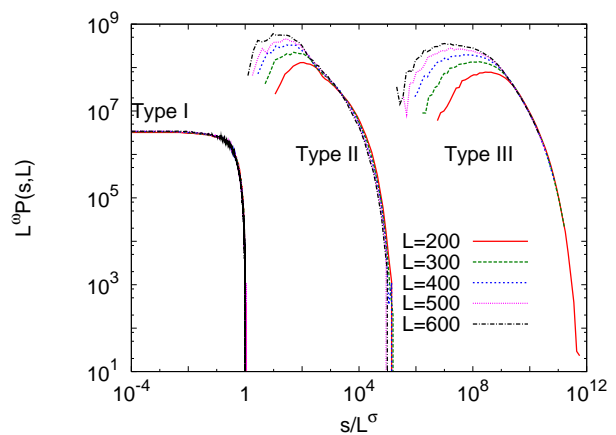}
\caption{(Color online) The scaled size distribution plots of the
three avalanches. The exponents for the type I, II and III
avalanches are $\sigma=1,2,3$, and $\omega=9/8, 11/4, 3$,
respectively. For a better view, the graphs for type
II and III have been
shifted by $10^{5}$ and $10^{12}$ units, on the x-axis.}
\label{fig:1ava}
\end{figure}

\begin{figure}
\includegraphics[scale=0.65,angle=0]{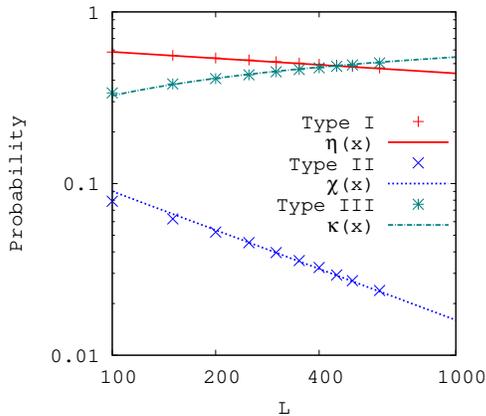}
\caption{(Color online) Probability of different types of avalanches. The fitting
functions are $\eta\left(x \right)=1.04/x^{1/8}$, $\chi\left( x
\right)=2.85/x^{3/4}$ and $\kappa\left( x \right)=1-\eta\left( x
\right)-\chi\left( x \right)$, for type I, II and III, respectively. }
\label{fig:relprob}
\end{figure}

We have analyzed our Monte-Carlo simulation results for five different
values of $L$, each averaged over $10^{6}$ trials.
The best data collapse of the numerical results is obtained for the values
$\alpha\simeq9/8$, $\beta\simeq11/4$ and $\nu\simeq3$ (Fig. \ref{fig:1ava}).
This implies that the probability of type I and type II avalanches
decreases as $L^{-1/8}$ and $L^{-3/4}$, for large $L$. This is consistent with the
numerical results in Fig. \ref{fig:relprob}. Thus, in the
thermodynamic limit, there will be mostly
type III avalanches.

The geometry of the lattice is important
in deciding the criticality of the steady state. To see this, we consider
a straight-forward generalization of our model on to other lattices.
First we consider a three-coordinated Caley tree.
Using an argument similar to the one-dimensional chain, one can see that all the stable configurations are equally probable in the steady state.
Then using the tree structure of the lattice, it is easy to show 
that the probability of an avalanche in which $r$ \textit{distinct} water
molecules rotate, has a form $2^{r}\exp\left( -k r/\log v \right)$, where $v$ is the
number of vertices on the Caley tree and $k$ is a numerical constant.
Then the steady state is not critical. On any other three-coordinated
lattice,
e.g., hexagonal lattice, there will be lesser freedom for the water
molecules. However, on lattices like square or
Kagome lattice, where each site is connected to four neighbors, the probability
distribution has a power-law tail, implying that the steady state is
critical.
On a lattice with coordination number greater than four, e.g., cubic
lattice, the water molecules become stable more easily, and the
probability of getting large avalanches
decreases exponentially. 

In an ice crystal, the water molecules are arranged on a four-coordinated diamond-like lattice.
Such a lattice is more like the Kagome lattice, and
the steady state of the icepile model is expected to be
critical. Although the dynamics of the model in its present form is not
very realistic (a more careful modeling of the intra-molecular
interaction and the dynamics is required), it does show an important qualitative
property of ice -- there is a long-range correlation in the hydrogen-bonded
network \cite{lrcexpt}.

In the Grotthuss mechanism, because the molecular reorientation is
slower than the $\textrm{H}^{+}$ hopping, the effective mobility of $\textrm{H}^{+}$ is
limited by that of the defects. However, when the defect-mobility was
estimated \cite{agmon} using experimental value of water rotation time \cite{nakahara} and assuming
a \textit{simple diffusion}, it was found that it is much less compared to
the experimental value of $\textrm{H}^{+}$ mobility, at low
temperatures.
This discrepancy can be resolved qualitatively, staying within the
Grotthuss formalism, if
one considers the long-range correlation in the hydrogen-bonded network. The correlated walk on the network
increases the mobility of the defects significantly, and
this in turn increases the mobility of $\textrm{H}^{+}$.

In \textit{summary}, we have proposed a dynamical model of the hydrogen-bonded
network in ice, the steady
state of which, in the slow driving limit, has a long-range correlation.
Ignoring this correlation in the study of proton conduction in real
ice,
can be misleading.
The model is also an example of SOC with non-conservative dynamics. Another important
aspect of the model is that its dynamics
is non-abelian, \textit{i.e.}, the final configuration depends on the order in which
the system is driven. All the known undirected models of SOC with exact solutions
are abelian \cite{dhar06}. There are only a handful of analytical results available for the
non-abelian models \cite{fey,lee}, but those are all in the special
limit, where the
dynamics essentially becomes abelian. In fact, even the one-dimensional
non-abelian models also are highly non-trivial. We solved
the icepile model in one dimension, and determined its steady state.
We also calculated the exact number of recurrent configurations
on a square lattice. A similar calculation is possible
on a Kagome lattice, and it can be shown that the number of recurrent
configurations
\begin{equation}
\Omega_{Kagome}=L!^{2}\times 4^{2L}(2L+1)^{L},
\end{equation}
on a $2L\times2L$ Kagome lattice lattice with open boundaries. 

\acknowledgments
I thank Deepak Dhar for introducing the central idea of
the work and his critical suggestions in the process of preparing
the manuscript. The idea of this model was originated from a computer
game, first brought to our attention by James Propp. I also
acknowledge Shaista Ahmad for her thorough and meticulous proofreading
of the manuscript.

\end{document}